\documentclass[twocolumn,showpacs,aps,prl,superscriptaddress,letterpaper]{revtex4}

\usepackage{graphicx}
\usepackage{dcolumn}
\usepackage{amsmath}
\usepackage{epsfig}
\usepackage{multirow}
\RequirePackage{xspace}
\usepackage{relsize}

\def\Imm       {\ensuremath{\Im m}}
\def\Ree       {\ensuremath{\Re e}}

\def\barpk{{\raise.35ex\hbox
{${\sss (}$}}--{\raise.35ex\hbox{${\sss )}$}}}
\def\bbarp{\hbox{$B$\kern-0.9em\raise1.4ex\hbox{\barpk}}}
\def\beq{\begin{equation}}
\def\eeq{\end{equation}}
\def\bea{\begin{eqnarray}}
\def\eea{\end{eqnarray}}
\def\nn{\nonumber}
\def\sss{\scriptscriptstyle}
\def\roughly#1{\mathrel{\raise.3ex\hbox
{$#1$\kern-.75em\lower1ex\hbox{$\sim$}}}}

\def\bd{B_d^0}
\def\bs{B_s}

\def\btos{{\bar b} \to {\bar s}}

\def\bra#1{\left\langle #1\right|}
\def\ket#1{\left| #1\right\rangle}
\def\fT{f_{\sss T}}
\def\fL{f_{\sss L}}
\def\fTfL{f_{\sss T}/f_{\sss L}}
\def \Kbar{\bar K}

\def\Abar{{\bar A}}

\def\pewp{P'_{\sss EW}}
\def\pewcp{P_{\sss EW}^{\sss \prime C}}

\def\L{{\cal L}}

\def\L{\Lambda_{\sss QCD}}

\def\Abar {\kern 0.18em\overline{\kern -0.18em A}{}\xspace}
\def\Kbar {\kern 0.18em\overline{\kern -0.18em K}{}\xspace}
\def\Bbar {\kern 0.18em\overline{\kern -0.18em B}{}\xspace}
\def\Dbar {\kern 0.18em\overline{\kern -0.18em D}{}\xspace}

\def\ve{\varepsilon}

\def\non{\nonumber}

\begin{document}

\begin{flushleft}
UMISS-HEP-2007-04 \\
UdeM-GPP-TH-07-162 \\[10mm]
\end{flushleft}

\title{
\large \bfseries \boldmath
 Study of Polarization in $B\to VT$ Decays
}

%
\author{Alakabha Datta}\thanks{datta@phy.olemiss.edu}
\affiliation{Dept.\ of Physics and Astronomy, 108 Lewis Hall,
University of Mississippi, Oxford, MS 38677-1848, USA}
\author{Yanyan Gao}\thanks{ygao@pha.jhu.edu}
\author{Andrei V. Gritsan}\thanks{gritsan@jhu.edu}
\affiliation{Dept.\ of Physics and Astronomy, Johns Hopkins
University, Baltimore, MD 21218, USA}
\author{David London}\thanks{london@lps.umontreal.ca}
\author{Makiko Nagashima}\thanks{makiko@lps.umontreal.ca}
\author{Alejandro Szynkman}\thanks{szynkman@lps.umontreal.ca}
\affiliation{Physique des Particules, Universit\'e de
Montr\'eal,C.P. 6128, succ. centre-ville, Montr\'eal, QC,
Canada H3C 3J7}

\date{\today}

\begin{abstract}
In this paper, we examine $B \to VT$ decays ($V$ is a vector
and $T$ is a tensor meson), whose final-state particles can
have transverse or longitudinal polarization. Measurements
have been made of $B\to\phi K_2^*$, and it is found that
$\fTfL$ is small, where $\fT$ ($\fL$) is the fraction of
transverse (longitudinal) decays.  We find that the standard
model (SM) naively predicts that $\fTfL \ll 1$.  The two
extensions of the naive SM which have been proposed to
explain the large $\fTfL$ in $B\to\phi K^*$ -- penguin
annihilation and rescattering -- make no firm predictions for
the polarization in $B\to\phi K_2^*$.  The two new-physics
scenarios, which explain the data in $B\to\pi K$ and the
$\phi( \rho) K^*$ polarization measurements, can reproduce
the $\fTfL$ data in $B\to\phi K_2^*$ only if the $B\to T$
form factors obey a certain hierarchy.  Finally, we present
the general angular analysis which can be used to get
helicity information using two- and three-body decays.
\end{abstract}


\maketitle


\section{Introduction}
An interesting effect has been observed in some $B\to V_1V_2$
decays ($V_i$ is a light charmless vector meson) which are
dominated by $\btos$ penguin transitions in the SM.  Because
the final-state particles are vector mesons, this decay is in
fact three separate decays, one for each polarization of the
vector mesons (one longitudinal, two transverse).  Naively,
the transverse amplitudes are suppressed by a factor of size
$m_{\sss V}/m_{\sss B}$ ($V$ is one of the vector mesons)
with respect to the longitudinal amplitude. As such, one
expects the fraction of transverse decays, $\fT$, to be much
less than the fraction of longitudinal decays, $\fL$.

The polarizations were first measured in $B\to\phi K^*$
decays \cite{phiK*}. There it was found that the two
fractions $\fT$ and $\fL$ are roughly equal: $\fTfL (B\to\phi
K^{*}) \simeq 1$. This was also seen in some $B\to\rho K^*$
decays \cite{belle:rhokst, babar:rhokst}. The latest data are
shown in Table~\ref{table1} \cite{belle:rhokst, babar:rhokst,
phiK*0, belle:phikst, cdf:phikst, babar:phikstpl,
babar:omegakst, belle:omegakst, babar:phikst3, hfag, pdg}.

\begingroup
\begin{table}[t!]
\caption{ Measurements of the branching fraction ${\cal B}$,
longitudinal polarization fraction $\fL$, and fraction of
parity-odd transverse amplitude $f_{\sss\perp}$ for $B\to\phi
K^*_{\sss J}$, $\rho K^*$, and $\omega K^*$ \cite{hfag, pdg}.
Numbers in parentheses indicate observables measured with
less than 4$\sigma$ significance.  For a complete list of up
to 12 parameters measured, including CP-violating
observables, see references quoted.  }
\begin{center}
{
\begin{tabular}{lccc}
\hline\hline
\vspace{-0.3cm}\\
 Mode & ${\cal B}$ (${ 10^{-6}}$) & $f_{\sss L}$  & $f_{\sss\perp}$ \\
\vspace{-0.3cm}\\
\hline
\vspace{-0.3cm}\\
 {$\phi K^{*}(892)^0$} \cite{phiK*0, belle:phikst, cdf:phikst} 
 & ${ 9.5}\pm{0.9}$ & ${ 0.49}\pm{0.03}$  & ${0.25}\pm 0.03$ \\
\vspace{-0.3cm}\\
 {$\phi K^{*}(892)^+$} \cite{belle:phikst, babar:phikstpl}
 & ${ 10.0}\pm 1.1$ & ${ 0.50}\pm 0.05$  & ${0.20}\pm0.05$ \\
\vspace{-0.3cm}\\
 $\rho^+ K^{*}(892)^{0}$ \cite{belle:rhokst, babar:rhokst} & ${ 9.2}\pm 1.5$ &  ${ 0.48}\pm 0.08$ \\
\vspace{-0.3cm}\\
 $\rho^0 K^{*}(892)^{0}$ \cite{babar:rhokst} & ${ 5.6}\pm{1.6}$  & ${ 0.57}\pm0.12$ \\
\vspace{-0.3cm}\\
 $\rho^- K^{*}(892)^{+}$ \cite{babar:rhokst} & {$<{ 12.0}$} ({$5.4^{+4.1}_{-3.8}$}) \\
\vspace{-0.3cm}\\
 $\rho^0 K^{*}(892)^{+}$ \cite{babar:rhokst} &  {$<{ 6.1}$}  ({$3.6^{+1.9}_{-1.8}$}) & $(0.9\pm 0.2)$ \\
\vspace{-0.3cm}\\
 $\omega K^{*}(892)^{0}$ \cite{babar:omegakst, belle:omegakst} & {$<{ 2.8}$} ({$1.6^{+0.8}_{-0.7}$}) &  \\
\vspace{-0.3cm}\\
 $\omega K^{*}(892)^{+}$ \cite{babar:omegakst} & $<{ 3.4}$ ($0.6^{+1.8}_{-1.5}$) &  \\
\vspace{-0.3cm}\\
  $\phi K^{*}(1680)^0$ \cite{babar:phikst3} & $<{3.5}$ &  \\
\vspace{-0.3cm}\\
 {$\phi K_2^{*}(1430)^0$} \cite{phiK*0} & ${7.8}\pm{1.3}$ & ${0.85}\pm 0.08$  & ${0.05}\pm0.05$\\
\vspace{-0.3cm}\\
  $\phi K_3^{*}(1780)^0$ \cite{babar:phikst3} &  $<{2.7}$ &  \\
\vspace{-0.3cm}\\
  $\phi K_4^{*}(2045)^0$ \cite{babar:phikst3} &  $<{15.3}$ &  \\
\vspace{-0.3cm}\\
\hline\hline
\end{tabular}
}
\end{center}
\label{table1}
\end{table}
\endgroup

The fact that there is a discrepancy between the observed
$\fTfL$ and the naive expectation could be a signal of
physics beyond the standard model (SM) \cite{NP}. Indeed, to
date, there have been several hints of such new physics (NP)
in $\btos$ transitions, though none has been statistically
significant. On the other hand, there are explanations within
the SM. Assuming that there is a single explanation for the
large $\fTfL$'s, there are two proposed SM solutions
\cite{BVV}: penguin annihilation \cite{Kagan} and
rescattering \cite{soni, SCET}.

However, one can also look at $B \to VT$ decays ($T$ is a
tensor meson). Here too there are three polarizations, and
$\fTfL$ can be measured. This has been done in $B\to\phi
K_2^*$ \cite{phiK*0}, and the results are shown in
Table~\ref{table1}. In this case $\fTfL (B\to\phi K_2^{*})$
is found to be small.

The various explanations must account for the $\fTfL$ data in
both $B\to V_1V_2$ and $B \to VT$ decays. In this paper we
examine this question, both in the SM and with NP. 

In Sec.~2, we look at the prediction of the naive SM, based
on factorization, for the polarizations in $B\to VT$ decays.
In Sec.~3, we examine how the expectations change for
$B\to\phi K_2^*$, $B\to\rho K_2^*$ and $B\to\omega K_2^*$ in
the presence of penguin annihilation or rescattering. In
Sec.~4, we discuss new-physics explanations/predictions for
polarizations in $B\to V_1V_2$ and $B \to VT$. Sec.~5
contains a discussion of the angular analysis to three-body
and two-body decays of the $B$-meson daughters. This is
relevant to $B \to VT$, and is applicable to the Chen-Geng
explanation \cite{ChenGeng} of polarizations in $B\to V_1V_2$
and $B \to VT$. We conclude in Sec.~6.

\section{Standard Model Prediction}

As detailed in the introduction, the SM naively predicts that
$\fTfL \ll 1$ in $B\to V_1V_2$ decays. In this section, we
examine the SM prediction for $\fTfL$ in $B\to VT$ decays.
As we will see, the analysis of $B\to VT$ decays is very
similar to that in $B\to V_1V_2$.

We begin by describing the kinematics involved in $B(p_{\sss
B}) \to V(q) T(p)$, where we have explicitly labeled the
momenta of the participating mesons. We work in the limit of
heavy mass for the initial hadron and large energy for the
final state \cite{formfactors}:
\beq 
(\L,m_{\sss V,T})\ll (m_{\sss B},E_{\sss V,T}) ~, 
\eeq
where $m_{\sss V,T}$ and $E_{\sss V,T}$ are the masses and
energies of the vector and tensor mesons, $m_{\sss B}$ is the
$B$ meson mass and $\L$ is the QCD scale. In the rest frame
of the $B$ we can therefore write
\bea
\label{v} 
p_{\sss B} &\equiv & m_{\sss B} (1,0,0,0) \non\\ 
p & \equiv & E_{\sss T} (1,0,0,1) \non\\ 
q & \equiv & E_{\sss V} ( 1,0,0,-1) ~.
\eea 
Here we have dropped terms of order $({m_{\sss V,T} / E_{\sss
V,T}})^2$ which is reasonable as $E_{\sss V,T} \sim m_{\sss
B}/2$.

We now specify the polarization vectors of the final-state
particles.  We define the polarization of the vector meson as
\begin{eqnarray}
\eta^{\mu}(0) & = & \frac{1}{m_{\sss V}}(q_z,0, 0, E_{\sss
V}) \approx \frac{1}{m_{\sss V}}(-E_{\sss V},0, 0, E_{\sss
V}) ~, \non\\
\eta^{\mu}(\mp) &= &\frac{1}{\sqrt{2}}(0, \mp 1, -i, 0) ~,
\end{eqnarray}
where we assume the vector meson is moving along the negative
z-axis. The polarization of the spin-2 tensor meson can be
specified by a symmetric and traceless tensor $s^{\mu\nu}$
which satisfies
\begin{eqnarray}
s^{\mu\nu}(p,h)&=& s^{\nu\mu}(p,h) ~, \non\\
s^{\mu\nu}(p,h)p_{\nu}&=&s^{\mu\nu}(p,h)p_{\mu}=0 ~,
\non\\
g_{\mu\nu}s^{\mu\nu} & = &0 ~,
\end{eqnarray}
where $h$ is the meson helicity.  The states of a massive
spin-2 particle can be constructed in terms of two spin-1
states as
\begin{eqnarray}
s^{\mu\nu}(\pm 2)&=&e^{\mu}(\pm) e^{\nu}(\pm) ~, \non \\
s^{\mu\nu}(\pm 1)&=&\frac{1}{\sqrt{2}}\left[ e^{\mu}(\pm)
  e^{\nu}(0)+e^{\mu}(0) e^{\nu}(\pm) \right] ~, \non \\
s^{\mu\nu}(0)&=& \frac{1}{\sqrt{6}}\left[
e^{\mu}(+)e^{\nu}(-) + e^{\mu}(-)e^{\nu}(+)\right] \non\\
&& ~~~+\sqrt{\frac{2}{3}}e^{\mu}(0) e^{\nu}(0)
\,,
\end{eqnarray}
where $e^{\mu}(0,\pm)$ denote the polarization vectors of a
massive vector state, and their explicit structures are
chosen as
\begin{eqnarray}
e^{\mu}(0) & = & \frac{1}{m_{\sss T}}(p_{\sss T},0, 0,
E_{\sss T}) ~,\non\\
e^{\mu}(\pm) &= &\frac{1}{\sqrt{2}}(0, \mp 1, -i, 0) ~,
\end{eqnarray}
where $m_{\sss T} (\vec{p}_{\sss T})$ is the mass (momentum)
of the particle and $E_T$ its energy. Since the $B$ meson is
a spinless particle, the helicities carried by decaying
particles in the two-body $B$ decay must be the same.  Thus,
although the tensor meson contains 5 spin degrees of freedom,
only $h=0$ and $\pm 1$ give nonzero contributions.  The $B
\to T$ transition form factors then involve the
``polarization'' vector $\ve_{\mu}$, defined as
\bea
 \ve_{\mu}(h) & = & s_{\mu\nu}(p,h)v^{\nu}(m_{\sss
 T}/p_{\sss T}) ~,
 \label{tpol}
 \eea
where $v$ is the velocity of the $B$ meson and $p_{\sss T}$
is the magnitude of the momentum of the tensor meson. In the
rest frame of the $B$ meson we have
 \bea
  \ve^{\mu}(\pm 2)&=&0 ~, \non\\
\ve^{\mu}(\pm 1) & = & \frac{1}{\sqrt{2}}\frac{m_{\sss
  T}}{p_{\sss T}}e(0)\cdot v e^{\mu}(\pm) ~, \non\\
\ve^{\mu}( 0)&=&\sqrt{\frac{2}{3}}\frac{m_{\sss T}}{p_{\sss T}}e(0)\cdot v e^{\mu}(0)\,,
 \eea
with $e(0)\cdot v=p_{\sss T}/m_{\sss T} \approx E_{\sss
T}/m_{\sss T}$.  Note that the polarization vector
$\ve_{\mu}(h)$ has the same energy scaling as the
polarization vector of a vector meson.  The structure of the
$B\to T$ form factors is the same as that of $B\to V $ with
$\ve_{\mu}(h)$ replacing the $V$ polarization vector.  

The next step involves the SM effective Hamiltonian for $B$
decays \cite{BuraseffH}:
\bea
H_{eff}^q &=& {G_F \over \protect \sqrt{2}}
[V_{fb}^*V_{fq}(c_1O_{1f}^q + c_2 O_{2f}^q) \nn\\
&& \hskip-2truecm - \sum_{i=3}^{10}(V_{ub}V^*_{uq} c_i^u
+V_{cb}V^*_{cq} c_i^c +V_{tb}V^*_{tq} c_i^t) O_i^q] + h.c.,
\label{Heff}
\eea
where the superscript $u$, $c$, $t$ indicates the quark which
is internal or involved in rescattering from the tree
diagram, $f$ can be the $u$ or $c$ quark, $q$ can be either a
$d$ or $s$ quark, and the $c_i$ are Wilson coefficients.  In
the following, we focus on the specific decay $B \to \phi
K_2^*$. The SM contribution to this process has a similar
structure to that in $B \to \phi K^*$ \cite{rparity, TP}:
\beq
A[B \to \phi K_2^*] = \frac{G_F}{\sqrt{2}}X P_{\phi} ~,
\eeq
with
\bea
X & = & - \sum_{q=u,c,t}V_{qb}^* V_{qs} \times \non\\
&& ~~~~
\left[
  a_3^q+a_4^q+a_5^q -\frac{1}{2}(a_7^q+a_9^q+a_{10}^q)
  \right] ~, \nn\\ 
P_{\phi} & = & m_{\phi}g_{\phi}\eta^{*\mu}_{\phi} \bra{K_2^*}
\bar{b} \gamma_{\mu}(1-\gamma_5)s \ket{B}, \
\label{BVT}
\eea
where 
\beq
a_i = 
\begin{cases}
c_i + {c_{i-1} / N_c} ~, & \text{$i$ even ~,} \\
c_i + {c_{i+1} / N_c} ~, & \text{$i$ odd ~.}
\end{cases}
\eeq
The quantities $m_{\phi}$, $g_{\phi}$ and
$\eta^{*\mu}_{\phi}$ represent the mass, decay constant and
the polarization four-vector of the $\phi$ meson. The various
form factors and decay constants are defined as
\bea
\bra{\phi} {\bar s} \gamma^\mu s \ket{0} & \!\! = \!\! &
g_{\sss \phi} m_{\sss \phi} \eta^{*\mu} ~, \nn\\
\bra{K_2^*} {\bar b} \gamma^\mu (1 \pm \gamma_5) s \ket{B}
\eta^{*}_\mu & = & \non\\
&& \hskip-2,2truecm {2 i \over m_{\sss B} + m_{\sss K_2^*} }
\widetilde{V} \epsilon^{\mu\nu\alpha\beta} p_\mu q_\nu
\epsilon^{*}_\alpha \eta^{* }_\beta \non\\
&& \hskip-2.2truecm ~\pm \left( m_{\sss B} + m_{\sss K_2^*}
\right) \widetilde{A}_1 \epsilon^{*} \cdot \eta^{*} \nn\\
&& \hskip-2.2truecm ~\mp~ \widetilde{A}_2 {2 \over m_{\sss B}
+ m_{\sss K_2^*} } \left( p \cdot \eta^{*}\right)\left( q \cdot
\epsilon^{*} \right) ~.
\label{FFT}
\eea
The form factors defined above are easily related to those
defined in Refs.~\cite{Isgur,Scora} using the definition in
Eq.~(\ref{tpol}).

We can now write down the various polarization amplitudes
from Eq.~(\ref{BVT}).  Using the matrix elements given above,
the polarization amplitudes are given by \cite{TP}
\bea 
A_{0} & \approx & { G_{\sss F} \over \sqrt{2}} 2 m_{\sss B} m_{\phi}
g_{\phi} X \sqrt{\frac{2}{3}}\left[ \left( \widetilde{A}_1 - \widetilde{A}_2 \right)\right.\nn\\
& & ~~~~~ +~\left.
\frac{m_{\sss K_2^*}}{m_{\sss B}} \left( \widetilde{A}_1 + \widetilde{A}_2 \right)
\right] \frac{m_{\sss B}^2}{4 m_{\phi} m_{\sss K_2^*}} ~, \nn\\
A_\| & \approx & - { G_{\sss F} \over \sqrt{2}} \sqrt{2} m_{\sss B}\frac{1}{\sqrt{2}}
\left[ m_{\phi} g_{\phi} \left( 1 + {m_{\sss K_2^*}\over m_{\sss B}} \right)
\widetilde{A}_1 X \right] ~, \nn\\
A_\perp & \approx & - { G_{\sss F} \over \sqrt{2}} \sqrt{2} m_{\sss B}\frac{1}{\sqrt{2}}
\left[ m_{\phi} g_{\phi} \left( 1 - {m_{\sss K_2^*}\over m_{\sss B}} \right)
\widetilde{V} X \right]~.
\label{pol}
\eea

In the large-energy limit, the tensor and vector form factors
are expressible in terms of two universal form factors
\cite{formfactors}. This is due to the simplified structure
for the various currents in the effective theory
\cite{formfactors}, which takes the form
\bea
\overline{q}_n b_v&=&v_\mu \overline{q}_n\gamma^\mu b_v\,,\label{courantsDeb}\\
\overline{q}_n\gamma^\mu b_v&=&n^\mu \overline{q}_nb_v+i\epsilon^{\mu\nu\rho\sigma}v_\nu n_\rho\overline{q}_n\gamma_\sigma\gamma_5b_v\,,\\
\overline{q}_n\gamma^\mu\gamma_5 b_v&=&-n^\mu \overline{q}_n\gamma_5b_v+i\epsilon^{\mu\nu\rho\sigma}v_\nu n_\rho\overline{q}_n\gamma_\sigma b_v\,,\\
\overline{q}_n\sigma^{\mu\nu}b_v&=&i\left[n^\mu v^\nu\overline{q}_n b_v
-n^\mu \overline{q}_n\gamma^\nu b_v-(\mu \leftrightarrow \nu)\right]
\non\\
&+& \epsilon^{\mu\nu\rho\sigma}v_\rho n_\sigma\overline{q}_n\gamma_5 b_v\,,\\
\overline{q}_n\sigma^{\mu\nu}\gamma_5 b_v&=&i\left[n^\mu v^\nu\overline{q}_n \gamma_5b_v
+n^\mu \overline{q}_n\gamma^\nu\gamma_5 b_v-(\mu \leftrightarrow \nu)\right]
\non\\
&+&\epsilon^{\mu\nu\rho\sigma}v_\rho n_\sigma\overline{q}_n
b_v\,.\label{currents}
\eea
The above relations are valid for both $B\to V$ and $B\to T$
form factors. Given this, and given the fact that the
transition form factors have the same structure, with proper
redefinition of the polarization vector for the tensor meson,
we expect that the $B\to T$ form factors should also be
expressible in terms of two universal form factors in the
large-energy limit.

Hence, in the large-energy effective theory (LEET), ignoring
possible power-suppressed and $\alpha_s$ corrections, we
have
\bea
 \widetilde{A}_{1} & \approx & \widetilde{\zeta}_\perp \left(
 1-{m_{\sss K_2^*}\over m_{\sss B}} \right) ~, \non\\
\widetilde{ A}_{2} & \approx &
\widetilde{\zeta}_\perp \left( 1+{m_{\sss  K_2^*}\over m_{\sss B}}\right) - 
{2 m_{\sss  K_2^*} \over m_{\sss B}} \widetilde{\zeta_\|} ~, \non\\
\widetilde{V}_{1}  & \approx & \widetilde{\zeta}_\perp\left(
 1+{m_{\sss  K_2^*}\over m_{\sss B}}\right) ~,
\label{ffrel}
\eea
where $\widetilde{\zeta}_\perp$ and $\widetilde{\zeta}_\|$
are the two universal form factors. Note that in the
effective theory there exists no relation between
$\widetilde{\zeta}_\perp$ and $\widetilde{\zeta}_\|$.  For
$B\to V$ form factors, most models find
$\widetilde{\zeta}_\perp$ and $\widetilde{\zeta}_\|$ to be of
similar size, but this may not be true for the $B\to T$ form
factors. However, there is not much literature on the
calculation of $B\to T$ form factors, and often the model
predictions are in large disagreement with each other
\cite{Scora, Hwang}. The model of Ref.~\cite{Scora} is not
expected to be reliable in the low-$q^2$ region and predates
the form-factor relations obtained in
Ref.~\cite{formfactors}. The model of Ref.~\cite{Hwang} has
form factors very different from those in Ref.~\cite{Scora}
and appears to be inconsistent with the form-factor relations
in LEET.

Given this, we will employ a general analysis assuming only
the form-factor relations from LEET. We will consider two
cases described below:

Case (a): We assume $\widetilde{\zeta}_\perp \approx
\widetilde{\zeta}_\|$. With the help of Eqs.~(\ref{pol}) and
(\ref{ffrel}) we find that
\bea
\frac{A_{\sss T}}{A_0} & \sim & \frac{m_\phi}{m_{\sss B}}
~,~~ \quad T= \perp, \| ~, \non\\
\frac{A_\perp}{A_{\|}} & \approx & 1 ~.
\label{a}
\eea
This follows from the fact that for this case
\bea
\widetilde{A}_2 &=& \widetilde{A}_1 +O(m_{\sss K_2^*}/m_{\sss B})
~, \non\\
\widetilde{V} &=& \widetilde{A}_1 +O(m_{\sss K_2^*}/m_{\sss B})
~.
\eea

Case (b): We assume $\widetilde{\zeta}_\perp \ll
\widetilde{\zeta}_\|$ with $\widetilde{\zeta}_\| \sim
({m_{\sss B} / m_{\sss K_2^*}})\widetilde{\zeta}_\perp $. In
this case, even though $\widetilde {A}_1$ and $\widetilde{V}$
differ only by terms of $O(m_{\sss K_2^*}/m_{\sss B})$, the
form factor $\widetilde{A}_2$ can be very different.
Consequently, with the help of Eqs.~(\ref{pol}) and
(\ref{ffrel}), we find that
\bea
\frac{A_{\sss T}}{A_0} & \sim & \frac{m_\phi
m_{\sss K_2^*}}{m_{\sss B}^2} ~,~~ \quad T= \perp, \| ~, ~\non\\
\frac{A_\perp}{A_{\|}} & \approx & 1 ~.
\label{b}
\eea
In both cases the longitudinal polarization dominates and the
transverse polarizations are of the same size, so that
$\fTfL$ is small. Although the decay $B\to\phi K_2^*$ was
used to derive this result, it holds for all $VT$ final
states. We therefore conclude that the SM predicts that
$\fTfL \ll 1$ in $B\to VT$ decays.

Finally, we can consider a third case, case (c), in which
$\widetilde{\zeta}_\| \ll \widetilde{\zeta}_{\perp}$ with
$\widetilde{\zeta}_\perp \sim ({m_{\sss B} / m_{\sss K_2^*}
}) \widetilde{\zeta}_\| $.  However, it is clear from
Eqs.~(\ref{pol}) and (\ref{ffrel}) that one obtains $A_{\sss
T}/A_0 \sim (m_\phi/m_{\sss B})
(\widetilde{\zeta}_\perp/\widetilde{\zeta}_\|) \sim
(m_\phi/m_{\sss K_2^*})$, so that $\fTfL \simeq 1$.  This is
in contradiction with the experimental results for $B \to
\phi K_2^*$ (Table~\ref{table1}).

Note that, in the presence of new physics, the predictions of
the three cases can be altered. Two sections below we turn to
the effect of NP.

\section{Penguin Annihilation and Rescattering}

Earlier, it was noted that the SM (naively) predicts that
$\fTfL \ll 1$ in $B\to V_1V_2$ decays, in contrast to
experimental results. It was also noted that there are
effects within the SM -- penguin annihilation \cite{Kagan}
or rescattering \cite{soni, SCET} -- that, if large, could
explain the observed value of $\fTfL$. In this section, we
review the action of penguin annihilation and rescattering in
$B\to V_1V_2$ decays, and establish their prediction for
$\fTfL$ in $B\to VT$ decays, specifically $B\to\phi K_2^*$,
$B\to\rho K_2^*$ and $B\to\omega K_2^*$.

\subsection{\boldmath $B\to\phi K_2^*$}

We begin by examining penguin annihilation (Fig.~1).
$B\to\phi K^*$ receives penguin contributions, ${\bar b}
{\cal O} s {\bar q} {\cal O} q$, where $q=u,d$ (${\cal O}$
are Lorentz structures, and color indices are
suppressed). Applying a Fierz transformation, these operators
can be written as ${\bar b} {\cal O}' q {\bar q} {\cal O}' s
$. A gluon can now be emitted from one of the quarks in the
operators which can then produce an $s\bar{s}$ quark
pair. These then combine with the ${\bar s}, q$ quarks to
form the final states $\phi K^{*+}~(q=u)$ or $\phi
K^{*0}~(q=d)$.

\begin{figure}[htbp]
\begin{center}
\begin{tabular}{cc}
\setlength{\epsfxsize}{0.98\linewidth}\leavevmode\epsfbox{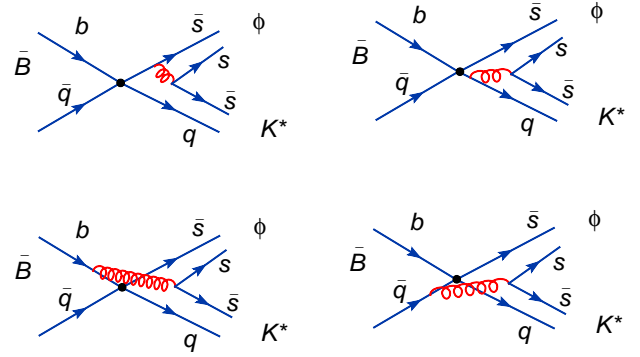}
\end{tabular}
\caption{The penguin annihilation diagrams.} 
\label{fig:narrow1}
\end{center}
\end{figure}

All annihilation contributions are usually expected to be
small as they are higher order in the $1/m_b$ expansion, and
thus ignored.  However, within QCD factorization (QCDf)
\cite{BBNS}, it is plausible that the coefficients of these
terms are large \cite{Kagan}. In QCDf, penguin annihilation
is not calculable because of divergences which are
parameterized in terms of unknown quantities. One may choose
these parameters to fit the polarization data in $B\to\phi
K^*$ decays.  (Within perturbative QCD (pQCD) \cite{pQCD},
the penguin annihilation is calculable and can be large,
though it is not large enough to explain the polarization
data in $B\to\phi K^*$ \cite{LM}.)

The second explanation of $\fTfL$ in $B\to\phi K^*$ is
rescattering (Fig.~2). It has been suggested that
rescattering effects involving charm intermediate states,
generated by the operator ${\bar b} {\cal O}' c {\bar c}
{\cal O}' s$, can produce large transverse polarization in
$B\to\phi K^*$.  A particular realization of this scenario is
the following \cite{soni}.  Consider the decay $B^+ \to
D_s^{*+} {\bar D}^{*0}$ generated by the operator ${\bar b}
{\cal O}' c {\bar c} {\cal O}' s $. Since the final-state
vector mesons are heavy, the transverse polarization can be
large.  The state $D_s^{*+} {\bar D}^{*0}$ can now rescatter
to $\phi K^{*+}$. If the transverse polarization $T$ is not
reduced in the scattering process, this will lead to
$B^+\to\phi K^{*+}$ with large $\fTfL$. (A similar
rescattering effect can take place for $\bd\to\phi K^{*0}$.)

\begin{figure}[htbp]
\begin{center}
\begin{tabular}{cc}
\setlength{\epsfxsize}{0.98\linewidth}\leavevmode\epsfbox{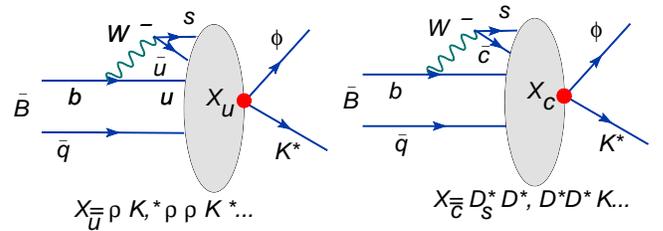}
\end{tabular}
\caption{The rescattering  diagrams.} 
\label{fig:narrow2}
\end{center}
\end{figure}

This concludes our review of explanations of $\fTfL$ in
$B\to\phi K^*$. The key question now is: what do these
predict for $\fTfL$ in $B\to\phi K_2^*$? In order to answer
this question, we must establish whether or not the
individual explanations depend on the final-state
particles. If they do not, then the prediction for $\fTfL$ in
$B\to\phi K_2^*$ will be the same as that in $B\to\phi K^*$,
which is in disagreement with experiment.

The calculation of penguin annihilation does depend on the
final-state wave function. Thus, it is possible to choose the
parameters such that $\fTfL$ is small in $B\to\phi K_2^*$ for
the three cases discussed above, in agreement with
experiment.

For rescattering, the same mechanism that takes place in
$B\to\phi K^*$ can occur in $B\to\phi K_2^*$: one requires
that $D_s^{*+} {\bar D}^{*0}$ rescatter to $\phi K_2^{*+}$.
However, any such rescattering has its own set of
parameters. It is again possible to choose parameters such
that $D_s^{*+} {\bar D}^{*0}$ rescattering to $\phi K_2^{*+}$
results in a small $\fTfL$ in $B\to\phi K_2^*$ for all the
three cases. Thus, rescattering can account for the small
$\fTfL$ result in $B\to\phi K_2^*$.

It is therefore possible that both explanations agree with
the $\fTfL$ measurements in $B\to\phi K^*$ and $B\to\phi
K_2^*$. However, this is not very satisfying, since there is
a new set of parameters for each final state, and it is
virtually impossible to calculate the values of the
parameters. We thus conclude that both penguin annihilation
and rescattering are viable, but not very convincing.

\subsection{\boldmath $B\to\rho K_2^*$}

Within the diagrammatic approach \cite{GHLR}, amplitudes for
$\btos$ processes can be written in terms of eight diagrams:
the color-favored and color-suppressed tree amplitudes $T'$
and $C'$, the gluonic penguin amplitudes $P'$, the
color-favored and color-suppressed electroweak penguin
amplitudes $\pewp$ and $\pewcp$, the annihilation and
exchange amplitudes $A'$ and $E'$, and the
penguin-annihilation diagram $PA'$. (The primes on the
amplitudes indicate $\btos$ transitions.) There are also
other diagrams, but they are much smaller and in general can
be neglected. Note that $P'$ includes rescattering
contributions from tree-level operators with up- and
charm-quark intermediate states.  However, using the
unitarity of the Cabibbo-Kobayashi-Maskawa (CKM) matrix, one
can write
\bea
P' & = & V_{ub}^* V_{us} P'_u + V_{cb}^* V_{cs} P'_c
+ V_{tb}^* V_{ts} P'_t \nn\\
& = & V_{ub}^* V_{us} (P'_u - P'_c) + V_{tb}^* V_{ts} (P'_t - P'_c)~.  
\eea
Thus, $P'$ includes the two quark amplitudes $P'_{tc} \equiv
(P'_t - P'_c)$ and $P'_{uc} \equiv (P'_u - P'_c)$. 

In principle, $\pewp$, $\pewcp$ and $PA'$ also receive
contributions from the $u$ and $c$ quarks. However, these are
negligible -- the three diagrams are dominated by the
intermediate $t$ quark. Hence all $\btos$ $B$-decay
amplitudes can be written in terms of the nine diagrams $T'$,
$C'$, $P'_{tc}$, $P'_{uc}$, $\pewp$, $\pewcp$, $A'$, $E'$ and
$PA'$.

From here on, we will redefine all nine diagrams by taking
them to have absorbed their associated CKM matrix elements.
In Ref.~\cite{GHLR}, the relative sizes of the (redefined)
amplitudes were estimated to be roughly
\bea
1 : |P'_{tc}| ~~,~~~~ {\cal O}({\bar\lambda}) : |T'|,~|\pewp|
~~ \nn\\
{\cal O}({\bar\lambda}^2) :
|C'|,~|P'_{uc}|,~|\pewcp|,~|PA'_{tc}| \nn\\
{\cal O}({\bar\lambda}^3) : |A'|, |E'|, ~~~~~~~~~
\label{hierarchy}
\eea
where ${\bar\lambda} \sim 0.2$. These SM estimates are often
used as a guide to neglect diagrammatic amplitudes and reduce
the number of parameters.

Now, $B\to\rho K_2^*$ \cite{Kim} actually represents four
processes. $B^+ \to \rho^+ K_2^{*0}$ is governed by the
underlying quark transition $\btos d{\bar d}$; $\bd \to
\rho^- K_2^{*+}$ is dominated by $\btos u{\bar u}$; $\bd \to
\rho^0 K_2^{*0}$ and $B^+ \to \rho^0 K_2^{*+}$ have both
transitions. Neglecting all diagrams of $ \le {\cal
O}({\bar\lambda}^2)$ [Eq.~(\ref{hierarchy})], the amplitudes
for these decays are given by
\bea
A(B^+ \to \rho^+ K_2^{*0}) & = & P'_{tc} ~, \nn\\
\sqrt{2} A(\bd \to \rho^0 K_2^{*0}) & = & P'_{tc} - \pewp ~, \nn\\
\sqrt{2} A(B^+ \to \rho^0 K_2^{*+}) & = & - P'_{tc} - T' \, e^{i\gamma}
- \pewp ~, \nn\\
A(\bd \to \rho^- K_2^{*+}) & = & - P'_{tc} - T' \, e^{i\gamma} ~.
\eea
The weak phase information in the CKM matrix is
conventionally parameterized in terms of the unitarity
triangle, in which the interior CP-violating angles are known
as $\alpha$, $\beta$ and $\gamma$ \cite{pdg}. In the above
amplitudes, we have explicitly written the dependence on the
weak phase $\gamma$, but the diagrams contain strong phases.

In naive factorization, at the mesonic level, the penguin
($P'_{t}$) and tree ($T'$) contributions take the form
$\bra{K_2^*} {\bar q} {\cal O} s \ket{0} \bra{\rho} {\bar b}
{\cal O} q \ket{B}$. This vanishes, because one cannot
produce the tensor meson $K_2^*$ from the vacuum. Thus, the
branching ratios for $B^+ \to \rho^+ K_2^{*0}$ and $\bd \to
\rho^- K_2^{*+}$ arise only due to nonfactorizable effects,
and are small. (The branching ratios for the other two decays
are not small due to the presence of $\pewp$ in the
amplitudes.)  If rescattering from the tree-level operators
is perturbatively calculable, as in QCDf, the size (and
phase) of $P'_{tc}$ is changed, but not the Lorentz
structure. As such, the branching ratios for $B^+ \to \rho^+
K_2^{*0}$ and $\bd \to \rho^- K_2^{*+}$ will remain small.
(If there are long-distance incalculable rescattering effects
then it is possible that the branching ratios of these decays
may not be small.) In penguin annihilation there are
nonfactorizable contributions to $B^+ \to \rho^+ K_2^{*0}$
and $\bd \to \rho^- K_2^{*+}$ of the form $\bra{\rho K_2^*}
{\bar q} {\cal O} s {\bar b} {\cal O} q \ket{B}$, which is
nonzero. As a result, the branching ratios for $B^+ \to
\rho^+ K_2^{*0}$ and $\bd \to \rho^- K_2^{*+}$ may {\it not}
be small. The measurement of these quantities will allow us
to test if large penguin annihilation or large
nonperturbative rescattering effects are present in these
decays.

Turning to polarizations, within penguin annihilation and
rescattering, the prediction of $\fTfL$ for each of the four
decays $B\to\rho K_2^*$ is arbitrary, as it depends on the
unknown set of parameters for each of the final states. 

It is tempting to try to get information from $B\to\phi
K_2^*$ by using flavor SU(3) for the final states. However,
this does not work.  Consider only the contribution of the
gluonic penguin to $B\to\phi K_2^*$ and $B\to\rho K_2^*$.  As
detailed above, the matrix element of this operator vanishes
for $B\to\rho K_2^*$, but it does not for $B\to\phi K_2^*$.
Thus, these two decays are not in fact related by SU(3).

We therefore conclude that there are no firm predictions for
$\fTfL$ in any of the $B\to\rho K_2^*$ decays from penguin
annihilation and rescattering.

\subsection{\boldmath $B\to\omega K_2^*$}

The decay $B\to\omega K_2^*$ represents two processes: $\bd
\to \omega K_2^{*0}$ and $B^+ \to \omega K_2^{*+}$. The
amplitudes are given by
\bea
\sqrt{2} A(\bd \to \omega K_2^{*0}) & = & P'_{tc} + 2
P'_{tc,dir} + {1\over 3} \pewp ~, \\
\sqrt{2} A(B^+ \to \omega K_2^{*+}) & = & P'_{tc} + 2
P'_{tc,dir} + T' \, e^{i\gamma} + {1\over 3} \pewp ~. \nn
\eea
At the quark level, $\omega = (d {\bar d} + u {\bar u}) /
\sqrt{2}$. Since the $\omega$ is an isosinglet, the gluon can
decay directly to it (unlike the $\rho^0$). However, this
requires the exchange of two additional gluons to absorb
color factors and is expected to be somewhat suppressed
according to the OZI rule. We denote this diagram by
$P'_{tc,dir}$ and note that this does not vanish in naive
factorization.

Due to the presence of $P'_{tc,dir}$ and $\pewp$ in the
amplitudes, the branching ratios for these decays may not
small in the SM. Still, it is interesting to note that the
branching fraction of the $B \to \omega K^*$ decays is
measured to be smaller than that of the corresponding $B \to
\rho K^*$ and $B \to \phi K^*$ decays, see
Table~\ref{table1}.  There is an intriguing enhancement of
the $B \to \omega (K\pi)^*$ decay \cite{belle:omegakst} with
a higher-mass $(K\pi)$ combination which makes it interesting
to compare $B \to VT$ and $B \to V_1V_2$ decays with
$\omega$, though no data on $B \to \omega K^*_2$ have been
presented yet. (There are no data on $B \to \rho K^*_2$ from
the $B$-factories either.)

As in the case of $B\to\rho K_2^*$, the prediction of $\fTfL$
for $B\to\omega K_2^*$ within penguin annihilation and
rescattering depends on the unknown set of parameters for
each decay, and is arbitrary.

\section{New Physics}

The CP measurements in many penguin decays that proceed
through $\btos$ transitions appear to be in conflict with
naive SM expectations. For example, the combined
branching-ratio and CP-asymmetry measurements in $B\to\pi K$
decays are at odds with the SM predictions \cite{piK}.
However, these measurements are not precise enough to draw
any firm conclusions about the existence of new physics. As
already noted, the polarization measurements in some $B\to
V_1V_2$ ($\btos$) decays also disagree with naive SM
estimates. It is therefore not unreasonable to attempt to
understand the data assuming NP. The important question to
ask is then the following: can we find a unified new-physics
explanation for all the discrepancies so far reported in
measurements of pure-penguin or penguin-dominated decays? An
attempt to answer this question was presented in
Ref.~\cite{NP}.  In this section we carry the analysis of
that paper to $B\to VT$ decays.

The basic philosophy of this approach is the following: we
assume that the naive SM predictions for the CP measurements
and polarizations are generally correct. In particular,
annihilation contributions are taken to be negligible, and we
assume that nonperturbative SM rescattering through charm
intermediate states is suppressed by $O(1/m_b)$. The
annihilation contributions are power suppressed in the
$1/m_b$ expansion and should be small. Indeed, there is no
clear experimental evidence of large annihilation
effects. The existence of large nonperturbative SM
rescattering, not suppressed by $O(1/m_b)$, is controversial
and highly model-dependent with very little predictive power.
In any case, a new-physics explanation of the polarization
data in $B \to V_1V_2$ decays is only called for if one
assumes that annihilation effects and nonperturbative SM
rescattering effects are power suppressed. We also neglect
rescattering from NP amplitudes. This can be justified by
explicit calculation of rescattering effects in QCDf and is
expected to be valid even in nonperturbative models of
rescattering \cite{NPrescatt}.

We begin with a general parametrization of NP as in
Ref.~\cite{NP}. It is
\beq  
{4 G_{\sss F} \over \sqrt{2}} \sum_{\sss A,B = L,R} \left\{ f_q^{\sss
AB} \, {\bar b} \gamma_{\sss A} s \, {\bar q} \gamma_{\sss B} q +
g_q^{\sss AB} \, {\bar b} \gamma^\mu \gamma_{\sss A} s \, {\bar q}
\gamma_\mu \gamma_{\sss B} q \right\} ~.
\label{npo}
\eeq
There are a total of 24 contributing operators ($A,B = L,R$,
$q=u,d,s$); tensor operators do not contribute to $B\to\pi
K$. For simplicity, we assume that a single operator
contributes, and we analyze their effects one by one. By
doing a fit to the $\pi K$ data and the $\rho K^*$
polarization measurements, the authors of Ref.~\cite{NP} were
able to conclude that operators with coefficients $f_d^{\sss
LL}$ or $f_d^{\sss RR}$ have to be present in any NP model to
explain the present data \cite{NP,piK}. Such operators may
easily arise in multi-Higgs models \cite{model23,bsmixing}.

The question now is whether these operators can explain the
polarization results in $\bd \to \phi K_2^*$.  However, at
the quark level, this decay takes the form $\btos s {\bar
s}$, i.e.\ it involves NP operators with the subscript `s'.
In order for $f_d^{\sss LL}$ or $f_d^{\sss RR}$ to be
relevant to $\bd \to \phi K_2^*$, it is necessary to assume
U-spin symmetry, i.e.\ $f_s^{\sss AB} = f_d^{\sss AB}$.

Let us now consider the process $\bd \to \phi K_2^*$, adding
the NP operator whose coefficient is $f_s^{\sss RR}$ ($=
f_d^{\sss RR}$) [Eq.~(\ref{npo})]:
\beq {4 G_{\sss F} \over \sqrt{2}} f_d^{\sss RR} \, {\bar b}
\gamma_{\sss R} s \, {\bar s} \gamma_{\sss R} s ~. \eeq
Because this is a scalar/pseudoscalar operator, within
factorization it may appear that this operator does not
contribute to $B_d \to \phi K_2^*$. However, it can affect
this process once we perform a Fierz transformation of this
operator (both fermions and colors):
\beq 
-{4 \over N_c} {G_{\sss F} \over \sqrt{2}} f_d^{\sss RR} \,
\left[ \frac12 \, {\bar b} \gamma_{\sss R} s \, {\bar s}
  \gamma_{\sss R} s - \frac18 \, {\bar b} \sigma^{\mu\nu}
  \gamma_{\sss R} s \, {\bar s} \sigma_{\mu\nu} \gamma_{\sss
    R} s \right] ~.
\label{tensor} \eeq
In order to estimate the effects of the NP operators on $B_d
\to \phi K_2^*$, we have to evaluate matrix elements of the
type
\bea
M_{\sss NP} & = & \langle \phi K_2^*|H_{\sss
  NP}|\bd\rangle  \sim  D_{\mu \nu} F^{\mu \nu} ~, \non\\
D_{\mu \nu} & = &\langle \phi| \bar{s} \sigma_{\mu \nu}
\gamma_{\sss R} s|0\rangle ~, \non\\
F^{\mu \nu} & =& \langle K_2^*| \bar{s} \sigma^{\mu \nu}
\gamma_{\sss R} b|\bd\rangle ~,
\eea
where $\sigma_{\mu \nu} \equiv (1/2)
[\gamma_\mu,\gamma_\nu]$.  We can calculate the NP matrix
elements as \cite{NP}
\bea & & Z_d^{\sss RR} \left\{ 2 \tilde{T}_2
\left( 1-
 { m_{\sss K_2^*}^2 \over m_{\sss B}^2 } \right)
 \left( \epsilon^* \cdot
\eta^* \right) \right. \non\\
& & \left. - {4 \over m_{\sss B}^2} \left( \tilde{T}_2
+ \tilde{T}_3 { m_{\sss \phi}^2 \over m_{\sss B}^2 } \right) \left(
\epsilon^* \cdot p_{\sss \phi} \right) \left( \eta^* \cdot p_{\sss K_2^*} \right)
 \right. \nn\\
& & \hskip2truecm \left. -~{4 i\over m_{\sss B}^2} \tilde{T}_1
\epsilon^{\mu\nu\alpha\beta} p_\mu q_\nu
\epsilon^{* }_\alpha \eta^{*}_\beta \right\} ~,
\eea
where the $\tilde{T}_i$ are form factors and
\beq 
Z_d^{\sss RR} \equiv -{1 \over 4 N_c} {G_{\sss F} \over
\sqrt{2}} f_d^{\sss RR} g_{\sss T}^{\sss \phi} m_{\sss B}^2
~.
\label{Zdef} 
\eeq
The various hadronic quantities are defined as
\bea
\bra{\phi} {\bar s} \sigma^{\mu\nu} s \ket{0} & \!\! =  \!\! & - i
\, g_{\sss T}^{\sss \phi} \left( \eta^{*\mu} q^\nu -
\eta^{*\nu} q^\mu \right) ~, \nn\\
\bra{K_2^*} {\bar b} \sigma^{\mu\nu} s \ket{B} q_\nu &
\!\! = \!\! & 
- 2 \tilde{T}_1 \epsilon^{\mu\nu\alpha\beta} q_\nu p_\alpha
\ve^{*}_\beta ~, \nn\\
\bra{K_2^*} {\bar b} \sigma^{\mu\nu} \gamma_5 s \ket{B} q_\nu  \!\!& =  \!\! &
  - i \tilde{T}_2 \left[ \left( m_{\sss B}^2 -
m_{\sss K_2^*}^2 \right) \epsilon^{* \mu} \right. \nn\\
& - & \left. \left( \epsilon^*
\cdot q
\right) \left( p_{\sss B}^\mu + p^\mu \right) \right] \nn\\
  &- &~i \tilde{T}_3 \left( \epsilon^* \cdot q \right) \left[ q^\mu \right. \non\\
 & & \hskip-6truemm \left. - { m_{\phi}^2 \over
m_{\sss B}^2
- m_{\sss K_2^*}^2} \left( p_{\sss B}^\mu + p_{\sss K_2^*}^\mu \right) \right], 
\eea
Working in the large-energy limit, we can then write
\bea
 \tilde{T}_{1}(q^2) & \approx & \tilde{\zeta}_\perp ~, \non\\
 \tilde{T}_{2}(q^2) & \approx & \tilde{\zeta}_\perp
\left ( 1-{ q^2 \over {m_{\sss B}^2-m_{\sss K_2^*}^2}} \right) ~, \non\\
 \tilde{T}_{3}(q^2) & \approx &
\tilde{\zeta}_\perp - 
{2 m_{\sss K_2^*} \over m_{\sss B}} \tilde{\zeta}_\| ~,
\eea
where $\tilde{\zeta}_\perp$ and $\tilde{\zeta}_\|$ are the
same two universal form factors that appear in the SM
predictions (Sec.~2). One can then write the various
polarization amplitudes as
\bea
A_0 & = & - 2 \sqrt{\frac{2}{3}} \, \tilde{\zeta}_\| {m_{\sss \phi} \over m_{\sss
B}} Z_d^{\sss RR} ~, \nn\\
A_\| & = & 2 \, \tilde{\zeta}_\perp Z_d^{\sss RR} ~,
\nn\\
A_\perp & = & 2 \, \tilde{\zeta}_\perp Z_d^{\sss RR} ~.
\eea 
This leads to
\bea
{A_{0} \over A_{\perp}} & = & -\sqrt{\frac{2}{3}} \, {m_{\sss \phi} \over m_{\sss B}}
{\tilde{\zeta}_\| \over \tilde{\zeta}_\perp} ~, \non\\
{A_{0} \over A_{\|}} & = & -\sqrt{\frac{2}{3}} \, {m_{\sss \phi} \over m_{\sss B}}
{\tilde{\zeta}_\| \over \tilde{\zeta}_\perp} ~, \non\\
{A_{\|} \over A_{\perp}} & = & 1 ~.
\label{np_pol}
\eea

It is therefore clear that the ratio of transverse to
longitudinal amplitudes depends on the value of the
form-factor ratio ${\tilde{\zeta}_\| / \tilde{\zeta}_\perp}$.
For the case (a) discussed in Sec.~2 with
$\widetilde{\zeta}_\perp \approx \widetilde{\zeta}_\|$, the
NP contribution to the longitudinal polarization is
suppressed relative to the transverse amplitudes. We then
have the prediction
\beq
\frac{A_{\sss T}}{A_0} = \frac{A_{\sss T}^{\sss SM}+A_{\sss
    T}^{\sss NP}}{A_0^{\sss SM}+A_0^{\sss NP}} \sim
    \frac{A_{\sss T}^{\sss NP}}{A_0^{\sss SM}} \sim O(1) ~,
\label{npcasea}
\eeq 
where $T= \perp, \|$.  Here we have used the fact that the NP
amplitude is of similar size to the SM amplitude
\cite{NP}. This prediction is consistent with data for the $B
\to \phi K^*$ decay but not for $B \to \phi K_2^*$.  Note
that for the $V_1V_2$ final state, most models do find the
universal form factors to be of similar size. Indeed,
Ref.~\cite{NP} finds that NP can explain $\fTfL$ in $B \to
\phi K^*$ with case (a). Hence either our assumption about
the values of the universal form factors for the $B \to V
K_2^*$ transition is wrong or we need a different kind of new
physics.

Let us now turn to case (b), which has
$\widetilde{\zeta}_\perp \ll \widetilde{\zeta}_\|$ with
$\widetilde{\zeta}_\| \sim ({m_{\sss B} / m_{\sss
V}})\widetilde{\zeta}_\perp $. Here, from Eq.~(\ref{np_pol}),
the NP contribution to $A_0$ is not suppressed and all NP
polarization amplitudes are of the same size. We then have
\bea
\frac{A_{\sss T}}{A_0} = \frac{A_{\sss T}^{\sss
SM}+A_{\sss T}^{\sss NP}}{A_0^{\sss SM}+A_0^{\sss NP}} 
& \sim & \frac{A_{\sss T}^{\sss NP}}{A_0^{\sss SM}} \non\\
& & \hskip-1truecm
\sim \frac{ (m_{\sss B}/m_\phi) A_{\sss T}^{\sss
              SM}} { (m_{\sss B}^2/m_{\phi} m_{\sss K_2^*})
              A_{\sss T}^{\sss SM}} \sim
              \frac{m_{\sss K_2^*}}{m_{\sss B}} ~.
\label{npcaseb}
\eea
This prediction is consistent with experiment. 

Finally, for case (c), which has $\widetilde{\zeta}_\| \ll
\widetilde{\zeta}_{\perp}$ the NP longitudinal amplitude is
very suppressed. Hence, in this case we obtain
\bea
\frac{A_{\sss T}}{A_0} = \frac{A_{\sss T}^{\sss
SM}+A_{\sss T}^{\sss NP}}{A_0^{\sss SM}+A_0^{\sss NP}} 
& \sim & \frac{A_{\sss T}^{\sss
SM}+A_{\sss T}^{\sss NP}}{A_0^{\sss SM}}\non\\ 
& \sim & 1 ~,
\eea
where we have assumed no cancellation between the SM and NP
transverse amplitudes. This case is, therefore, inconsistent
with experiment.

It appears that the NP scenario is the same as that of the SM
-- the prediction of $\fTfL$ in $B \to \phi K_2^*$ depends on
the values of unknown parameters.  Here it is
${\tilde{\zeta}_\| / \tilde{\zeta}_\perp}$. However, the
difference is that, with penguin annihilation and
rescattering, the parameters are essentially incalculable,
while the NP prediction depends on form factors.  Although
the values of these form factors are not very well known at
the moment, they can be calculated. We strongly urge that the
$B\to T$ form factors be computed.

It should be pointed out that an actual NP calculation of
$\fTfL$ in $B \to \phi K_2^*$ will require knowledge of the
form factors as well as the relative sizes of the NP and SM
amplitudes. However, we believe our naive estimate of the
polarization fractions will hold even when a detailed
calculation is carried out.

We now turn to $B \to \rho K_2^*$ decays. The decays $B^+ \to
\rho^+ K_2^{*0}$ and $\bd \to \rho^- K_2^{*+}$ are
particularly interesting because they vanish within naive
factorization in the SM. The NP contributions to these
amplitudes take the form
\bea
M_{\sss NP} & = & \langle \rho K_2^* |H_{\sss NP}|B\rangle
\sim D_{\mu \nu} F^{\mu \nu} ~, \non\\
D_{\mu \nu} & = &\langle K_2^*| \bar{d} \sigma_{\mu \nu}
\gamma_{\sss A} s|0\rangle ~, \non\\
F^{\mu \nu} & =& \langle \rho| \bar{b} \sigma^{\mu \nu}
\gamma_{\sss B} d|B \rangle ~.
\eea
However, the factor $D_{\mu \nu}=0$ as we cannot construct an
antisymmetric tensor out of the symmetric polarization tensor
or the momentum vector of the tensor meson. Hence there is no
NP contribution to $B^+ \to \rho^+ K_2^{*0}$ and $\bd \to
\rho^- K_2^{*+}$, and so they vanish within factorization.
It will be very interesting to measure these branching
ratios.

By the same logic, the NP does not affect the other two
decays, $\bd \to \rho^0 K_2^{*0}$ and $B^+ \to \rho^0
K_2^{*+}$, either.  However, this is less important since the
SM prediction for these branching ratios is not precise.

Finally, since the NP cannot affect any of the $B \to \rho
K_2^*$ decays, the prediction for $\fTfL$ here is the same as
that of the SM. Specifically, $\fTfL$ is expected to be small
in $\bd \to \rho^0 K_2^{*0}$ and $B^+ \to \rho^0 K_2^{*+}$ in naive factorization.
It will be important to measure the polarization in these
decays in order to test the SM and this type of NP.


\section{Angular Analysis}

In the previous sections, we have concentrated on the process
$B \to VT$ without paying attention to how the $T$ decays.
Here we focus on this issue. If the spin-2 $T$ has positive
parity, it decays principally to two pseudoscalars, e.g.\ $K
\pi$, though a three-pseudoscalar final state is also
possible, e.g.\ $K \pi\pi$.  For example, the branching
fraction of $K_2^*(1430)$ decay to $K\pi$ is roughly 1/2 and
to $K\pi\pi$ is roughly 1/3~\cite{pdg}.  However, if the
spin-2 $T$ has negative parity (or the spin-1 $A$ in $B \to
VA$ has positive parity -- $A$ is an axial-vector meson), its
decay to two pseudoscalars is forbidden and it decays
principally to three pseudoscalars.  These two cases of
two-body and three-body decays can be treated separately.

Indeed, recent experimental studies have concentrated on the
two-body decays of the $K^*$ meson in $B \to \phi
K^*$. However, this limits the study only to the $J^P = 1^-,
2^+$, etc.\ strange-meson states with $P=(-1)^{J}$, see
Table~\ref{table1}.  Here we point out that new information
can be obtained from polarization studies of the $B$-meson
decays to states with $J^P = 1^+, 2^-$, etc., i.e.\ with
$P=(-1)^{J+1}$, such as $K_1$ and $K_2$ mesons. An example of
such a decay is $B \to \phi K_{\sss J}$, and one needs to
reconstruct $K_{\sss J}$ final states with at least three
pseudoscalar mesons, such as $K_1 \to K \pi \pi$. The angular
distribution of the $B \to \phi K_{\sss J}$ decay products
becomes more complex and a full angular analysis requires a
new formalism.

\begin{figure}[t]
\begin{center}
\begin{tabular}{cc}
\setlength{\epsfxsize}{0.98\linewidth}\leavevmode\epsfbox{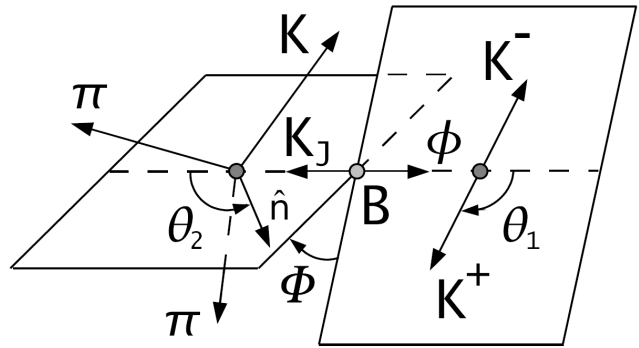}
\end{tabular}
\caption{ Definition of the helicity angles in the decay
$B\to\phi K_{\sss J}$ for the two-body ($\theta_1$) and
three-body ($\theta_2$) $B$-daughter decays, where both
angles are defined in the rest-frame of the decaying
meson. Normal to the three-body decay plane ($\hat{n}$) and
the daughter direction in the two-body decay serve as
analyzers of the polarization.  }
\label{fig:angles}
\end{center}
\end{figure}

The angular distribution of the $B\to V K^{(*)}_{\sss J}$
decay can be expressed as a function of $\theta_1$,
$\theta_2$, and $\Phi$, see Fig.~\ref{fig:angles}.  Here,
$\theta_1$ and $\theta_2$ are the helicity angles of the $V$
and the $K^{(*)}_{\sss J}$ resonances, defined as the angles
between the direction of the daughter meson (e.g. $K$ in
$\phi\to K\Kbar$) or normal to the three-body decay plane
(e.g. for $K_{\sss J}\to K\pi\pi$) and the direction opposite
the $B$ in the $V$ or $K^{(*)}_{\sss J}$ rest frame.  The
$\Phi$ is the angle between the decay planes of the two
systems, defined by the $B$ meson decay axis and the
direction of the daughter or normal as discussed above.

The analysis of the two-body angular distribution of the
particles has been widely used in polarization
measurements~\cite{jacob1959}, see also
Ref.~\cite{kramer1992} for application to $B\to V_1V_2$
decays.  It was pointed out in Ref.~\cite{berman1965} that in
the three-body decay of a particle, the normal to the decay
plane replaces the center-of-mass momentum as the analyzer of
the polarization.  There are more degrees of freedom in the
three-body decay, such as a Dalitz-plot structure. However,
those dynamical degrees of freedom can be integrated out and
simple results can be obtained from rotational and inversion
invariance.  We therefore proceed by deriving the angular
distributions of various two-body $B$ meson decays to mesons
with different spin and parity, which in turn decay strongly
to two or three pseudoscalar mesons. This will have direct
application to a number of polarization measurements
discussed in this paper.

We start by extending the angular formalism of a $B$-meson
decay (or any other spinless particle for that matter) to two
particles $X_1$ and $X_2$ with spins $J_1$ and $J_2$ and
parity $P_i=(-1)^{J_i}$. Each of the two particles decays
strongly to two pseudoscalars $X_1\to P_aP_b$ and $X_2\to
P_cP_d$\,, thus conserving parity.  Following the two-body
decay formalism \cite{jacob1959}, we obtain (see also
Refs.~\cite{phiK*0, babar:phikst3}):

\begin{widetext}
\begin{eqnarray}
\label{eq:helicity-twobody}
{1\over \Gamma}
{d^3\Gamma \over d{\cos\theta}_1 d{\cos\theta}_2d\Phi} =
{1\over \sum|A_{\lambda}|^2}
\left|~\sum_{\lambda}^{}
A_{\lambda} Y_{J_1}^{-\!\lambda}(\pi-\theta_1,-\Phi)
\,Y_{J_2}^{\lambda}(\theta_2,0)~\right|^2,
\end{eqnarray}
\end{widetext}
where $Y_{J}^{\lambda}$ are the spherical harmonics and the
sum is over the helicity values ($\lambda$ takes all discrete
values between $-j$ and $+j$, with $j$ being the smaller of
the two spins $J_1$ and $J_2$). The $A_{\lambda}$ is the
complex helicity amplitude in the $B$ decay, where $A_{0}$
corresponds to longitudinal polarization.  Examples of the
decays described by the above formula are $B\to \phi
K^*_{\sss J}$ and $B\to \rho K^*_{\sss J}$ with $J_1=1$ and
$J_2=J$.  In all of these cases we have three complex
amplitudes $A_\lambda$ and six real terms $\alpha_i$ which
appear in the angular distribution:

\begin{widetext}
\begin{eqnarray}
\alpha_1={|A_0|^2 \over \Sigma|A_\lambda|^2} = f_{\sss L} 
\label{eq:alpha1}
\end{eqnarray}
\begin{eqnarray}
\alpha_2
={|A_{\parallel}|^2 + |A_{\perp}|^2 \over \Sigma|A_\lambda|^2}
={|A_{+1}|^2 + |A_{-1}|^2 \over \Sigma|A_\lambda|^2}
=(1-f_{\sss L})
\label{eq:alpha2}
\end{eqnarray}
\begin{eqnarray}
\alpha_3
={ |A_{\parallel}|^2 - |A_{\perp}|^2  \over \Sigma|A_\lambda|^2}
={2}\cdot{{\Ree}(A_{+1}A^*_{-1}) \over \Sigma|A_\lambda|^2}
=(1-f_{\sss L}-2\cdot f_{\sss \perp})
\label{eq:alpha3}
\end{eqnarray}
\begin{eqnarray}
\alpha_4
={{\Imm}(A_{\perp}A^*_{\parallel})\over \Sigma|A_\lambda|^2}
={{\Imm}(A_{+1}A^*_{-1}) \over \Sigma|A_\lambda|^2}
=\sqrt{f_{\sss \perp}\cdot(1-f_{\sss L}-f_{\sss \perp})}\cdot\sin(\phi_{\sss \perp}-\phi_{\sss \parallel})
\label{eq:alpha4}
\end{eqnarray}
\begin{eqnarray}
\alpha_5
={{\Ree}(A_{\parallel}A^*_0)  \over \Sigma|A_\lambda|^2}
={ {\Ree}(A_{+1}A^*_0 + A_{-1}A^*_0) \over \sqrt{2}\cdot\Sigma|A_\lambda|^2}
=\sqrt{f_{\sss L}\cdot(1-f_{\sss L}-f_{\sss \perp})}\cdot\cos(\phi_{\sss \parallel})
\label{eq:alpha5}
\end{eqnarray}
\begin{eqnarray}
\alpha_6
={{\Imm}(A_{\perp}A^*_0)  \over \Sigma|A_\lambda|^2}
={ {\Imm}(A_{+1}A^*_0 - A_{-1}A^*_0) \over \sqrt{2}\cdot\Sigma|A_\lambda|^2}
=\sqrt{f_{\sss \perp}\cdot f_{\sss L}}\cdot\sin(\phi_{\sss \perp})
\label{eq:alpha6}
\\
\non
\end{eqnarray}
\end{widetext}

Here we use the definition adopted in the literature
\cite{pdg} with four convenient real terms:
\begin{eqnarray}
f_{\sss L}={|A_0|^2/\Sigma|A_\lambda|^2} ~, \\
f_{\sss \perp}={|A_{\perp}|^2/\Sigma|A_\lambda|^2}~, \\
\phi_{\sss \parallel} = {\rm arg}(A_{\parallel}/A_0) ~, \\
\phi_{\sss \perp} = {\rm arg}(A_{\perp}/A_0) ~.
\end{eqnarray}
We do not discuss CP violation in the angular distributions
here.

There are two special cases we discuss in more detail, namely
$J_1=1$ and $J_2=1$ or $2$, corresponding to $B\to V_1V_2$ or
$B\to VT$ decays.  From Eq.~(\ref{eq:helicity-twobody}), it
follows for $J_1=J_2=1$ (e.g. $B\to \phi K^*$, $\phi\to
K\Kbar$, $K^*\to K\pi$) that

\begin{eqnarray}
{{8\pi} \over {9\Gamma}}~\!{d^3\Gamma \over
d{\cos\theta}_1 d{\cos\theta}_2d\Phi }
=  \alpha_1\times \, \cos^2\theta_1\,\cos^2\theta_2 \non\\
+\,\alpha_2\times \, {1\over 4}\,\sin^2\theta_1\,\sin^2\theta_2 \non \\
+\,\alpha_3\times \, {1\over 4}\,\sin^2\theta_1\,\sin^2\theta_2 \,\cos2\Phi \non \\
-\,\alpha_4\times \, {1\over 2}\,\sin^2\theta_1\,\sin^2\theta_2 \,\sin2\Phi \non \\
+\,\alpha_5\times \, {1\over 2\sqrt{2}}\,\sin2\theta_1\,\sin2\theta_2\,\cos\Phi \non \\
-\,\alpha_6\times \, {1\over 2\sqrt{2}}\,\sin2\theta_1\,\sin2\theta_2\,\sin\Phi\, ,
\label{eq:vv-angular}
\end{eqnarray}
which is the well-known distribution for $B\to V_1V_2$
\cite{kramer1992}.  

From Eq.~(\ref{eq:helicity-twobody}), it follows that for
$J_1=1$ and $J_2=2$ (e.g.\ $B\to \phi K_2^*$, $\phi\to
K\Kbar$, $K_2^*\to K\pi$), the angular distribution is
described by

\begin{eqnarray}
{{32\pi} \over {15\Gamma}}~\!{d^3\Gamma \over
d{\cos\theta}_1 d{\cos\theta}_2d\Phi }
=  \alpha_1\times \, \cos^2\theta_1\, (3\,\cos^2\theta_2-1)^2 \non \\
+\,\alpha_2\times \, {3\over 4}\,\sin^2\theta_1\,\sin^2 2\theta_2 \non \\
+\,\alpha_3\times \, {3\over 4}\,\sin^2\theta_1\,\sin^2 2\theta_2\,\cos2\Phi \non \\
-\,\alpha_4\times \, {3\over 2}\,\sin^2\theta_1\,\sin^2 2\theta_2\,\sin2\Phi \non \\
+\,\alpha_5\times \, {\sqrt{6}\over2}\,\sin2\theta_1\,\sin2\theta_2\,(3\cos^2\theta_2-1)\,\cos\Phi \non \\
-\,\alpha_6\times \, {\sqrt{6}\over2}\,\sin2\theta_1\,\sin2\theta_2\,(3\cos^2\theta_2-1)\,\sin\Phi  \,.~~~
\label{eq:vt-angular}
\end{eqnarray}
The angular distribution for states with any other integer
value of $J_1$ and $J_2$ can be easily obtained from
Eq.~(\ref{eq:helicity-twobody}).

Next, we turn to the angular formalism of a spinless particle
decay to two particles with spins $J_1$ and $J_2$, $B\to X_1
X_2$, where the first particle decays to two pseudoscalars
$X_1\to P_aP_b$ and the second decays to three pseudoscalars
$X_2\to P_cP_dP_e$.  We assume parity conservation in the
$X_1$ and $X_2$ decays, i.e.\ $P_1=(-1)^{J_1}$. However,
there is no requirement on the parity $P_2$ of the second
particle.  We have an additional phenomenological amplitude
$F_m$ for the decay $X_2\to P_cP_dP_e$.  It depends on the
$X_2$ spin eigenvalue $m$ but not on $\lambda$.  Interference
between different $F_m$ amplitudes vanishes when integrated
over the rotation angle around the normal to the decay plane
and we are left with $2J_2+1$ real parameters $|R_m|^2$.
Additional symmetry considerations could put constraints on
$|R_m|^2$ values, as discussed below and in
Ref.~\cite{berman1965}.  Following the three-body decay
formalism \cite{berman1965}, we obtain

\begin{widetext}
\begin{eqnarray}
\label{eq:helicity-twoandthreebody}
{1\over \Gamma}
{d^3\Gamma \over d{\cos\theta}_1 d{\cos\theta}_2d\Phi} \propto
{1\over \sum|A_{\lambda}|^2}
\sum_{m}^{}|R_m|^2\,
\left|~\sum_{\lambda}^{} A_{\lambda} Y_{J_1}^{-\!\lambda}(\pi-\theta_1,-\Phi)
{\,d^{J_2\,}_{\lambda,m}}(\theta_2)~\right|^2 \, ,
\end{eqnarray}
\end{widetext}
where we omit the normalization factor for simplicity (in
general it would include a combination of the $|R_m|^2$
parameters).  The index $m$ runs from $-J_2$ to $+J_2$ and
$\lambda$ runs from $-j$ and $+j$ (with $j$ again being the
smaller of $J_1$ and $J_2$).

Additional constraints appear in
Eq.~(\ref{eq:helicity-twoandthreebody}) from parity
conservation in the $X_2\to P_cP_dP_e$ decay: for even (odd)
parity of $X_2$ only odd (even) values of $m$ contribute,
that is $P_2=(-1)^{m+1}$ \cite{berman1965}. This results in
only $m=0$ contributing to the decays with $J_2^{P_2}=1^-$,
such as $B\to K^*_{\sss J} \omega$, with $K^*_{\sss J}\to
K\pi$ and $\omega\to\pi^+\pi^-\pi^0$, and
Eq.~(\ref{eq:helicity-twoandthreebody}) reduces to
Eq.~(\ref{eq:helicity-twobody}) due to the simple
relationship of the $d^J_{\lambda,m}$ functions with $m=0$
and the spherical harmonics $Y^\lambda_J$.  It has been
pointed out earlier, e.g.\ in Ref.~\cite{kramer1992}, that
Eq.~(\ref{eq:vv-angular}) applies to $B\to V_1V_2$ decays
with $V_2\to P_cP_dP_e$, though it was incorrectly stated in
Ref.~\cite{kramer1992} that either one of the three daughter
momenta or the normal to the decay plane could define
$\theta_2$ and $\Phi$ in this formula. Consequently,
Eq.~(\ref{eq:helicity-twobody}) applies to the analysis of
the $B\to TV$ decay with $T\to P_aP_b$ and $V\to P_cP_dP_e$,
and in particular Eq.~(\ref{eq:vt-angular}) describes the
$B\to K^*_{2}\omega$ decay.

In a general case of $J_2^{P_2}\ne 1^-$ quantum numbers, more
than one $|R_m|^2$ parameter contributes.  In the following,
let us consider $J_1^{P_1}=1^-$ and $J_2^{P_2}=1^+, 2^+, 2^-$
states. Other final states could be considered by analogy. In
this case we still have three complex amplitudes $A_\lambda$,
but new real terms $\alpha_i$ may appear in the angular
distribution in addition to those shown in
Eqs.~(\ref{eq:alpha1}--\ref{eq:alpha6}):
\begin{widetext}
\begin{eqnarray}
\alpha_7
={{\Ree}(A_{\perp}A^*_\parallel)  \over \Sigma|A_\lambda|^2}
= {|A_{+1}|^2-|A_{-1}|^2  \over 2\cdot\Sigma|A_\lambda|^2}
=\sqrt{f_{\sss \perp}\cdot(1-f_{\sss L}-f_{\sss \perp})}\cdot\cos(\phi_{\sss \perp}-\phi_{\sss \parallel})
\label{eq:alpha7}
\\
\alpha_8
={{\Imm}(A_{\parallel}A^*_0)  \over \Sigma|A_\lambda|^2}
={ {\Imm}(A_{+1}A^*_0 + A_{-1}A^*_0) \over \sqrt{2}\cdot\Sigma|A_\lambda|^2}
=\sqrt{f_{\sss L}\cdot(1-f_{\sss L}-f_{\sss \perp})}\cdot\sin(\phi_{\sss \parallel})
\label{eq:alpha9}
\\
\alpha_9
={{\Ree}(A_{\perp}A^*_0)  \over \Sigma|A_\lambda|^2}
={ {\Ree}(A_{+1}A^*_0 - A_{-1}A^*_0) \over \sqrt{2}\cdot\Sigma|A_\lambda|^2}
=\sqrt{f_{\sss \perp}\cdot f_{\sss L}}\cdot\cos(\phi_{\sss \perp})
\label{eq:alpha8}
\\
\label{eq:helicitycoeffnew}
\non
\\ \non
\\ \non
\end{eqnarray}
\end{widetext}

Let us redefine some of the parameters:
\begin{eqnarray}
r_m \equiv { |R_m|^2 - |R_{-m}|^2 \over |R_m|^2 + |R_{-m}|^2 } \,; 
\label{eq:rm}
\\
r_{02} \equiv { |R_0|^2 \over |R_2|^2 + |R_{-2}|^2 } \, .
\label{eq:r02}
\end{eqnarray}

From Eq.~(\ref{eq:helicity-twoandthreebody}), it follows for
$J_1=1$, $J_2=2$, and $P_2=+1$ (e.g. $B\to \phi K_2^*$,
$\phi\to K\Kbar$, $K_2^*\to K\pi\pi$) that
\begin{eqnarray}
{{64\pi} \over {45\Gamma}}~\!{d^3\Gamma \over
d{\cos\theta}_1 d{\cos\theta}_2d\Phi }
=  \alpha_1\times \,\cos^2\theta_1\, \sin^2 2\theta_2 \non\\
+\,\alpha_2\times \,{1\over 3}\,\sin^2\theta_1\, (\cos^2\theta_2+\cos^2 2\theta_2) \non \\
-\,\alpha_3\times \,{1\over 3}\,\sin^2\theta_1\, (\cos^2\theta_2-\cos^2 2\theta_2) \,\cos2\Phi \non \\
+\,\alpha_4\times \, {2\over 3}\,\sin^2\theta_1\,(\cos^2\theta_2-\cos^2 2\theta_2) \,\sin2\Phi \non \\
-\,\alpha_5\times \,{1\over \sqrt{6}}\,\sin2\theta_1\, \sin4\theta_2\,\cos\Phi \non \\
+\,\alpha_6\times \,{1\over \sqrt{6}}\,\sin2\theta_1\, \sin4\theta_2\,\sin\Phi \non \\
+\,\alpha_7\times \, r_1\, {4\over 3}\,\sin^2\theta_1\,\cos\theta_2\,\cos2\theta_2  \non \\
+\,\alpha_8\times \, r_1\,\sqrt{2\over 3}\, \sin2\theta_1\,\cos\theta_2\,\sin2\theta_2\,\sin\Phi \non \\
-\,\alpha_9\times \, r_1\,\sqrt{2\over 3}\, \sin2\theta_1\,\cos\theta_2\,\sin2\theta_2\,\cos\Phi \, .
\label{eq:vt-3body-angular}
\end{eqnarray}

It is worth noting that three new angular terms $\alpha_7$,
$\alpha_8$, and $\alpha_9$ appear in
Eq.~(\ref{eq:vt-3body-angular}), and will appear below in
Eq.~(\ref{eq:va-angular}) together with the asymmetry term
$r_1$ from Eq.~(\ref{eq:rm}).  These terms would vanish if we
had a symmetry with respect to the inversion of the normal to
the decay plane for $X_2\to P_cP_dP_e$, that is between the
$m$ and $-m$ terms. As was pointed out in
Ref.~\cite{berman1965}, examples of such cases are two
identical pseudoscalar particles or pions in an eigenstate of
isotopic spin.  For example, $r_1=0$ for a sequential decay
like $K_1\to\rho K\to\pi\pi K$.  However, in the more general
case, $r_1$ is bound to $-1\le r_1\le +1$ and is a priori
unknown without the study of the $X_2\to P_cP_dP_e$ dynamics.

A nonzero value of $r_1$ may allow the determination of the
new angular terms in Eqs.~(\ref{eq:alpha7}--\ref{eq:alpha9}),
which would resolve discrete phase ambiguities for
$(\phi_\parallel,\phi_\perp)$, or, equivalently, the
ambiguity hierarchy of the $A_+$ and $A_-$ amplitudes.  For
any given values of $(\phi_\parallel,\phi_\perp)$, the simple
transformation $(-\phi_\parallel,\pi-\phi_\perp)$ preserves
the values of Eqs.~(\ref{eq:alpha1}--\ref{eq:alpha6}) and
leads to an ambiguity in experimental measurements. This
difficulty has been resolved for the final states with
$K^*(892)$ with the help of additional angular terms which
appear in the interference between the $K^*(892)$ and
$K^*_0(1430)$ \cite{phiK*0, babar:phikstpl, babar:jpsikst}.
However, the presence of nonzero terms in
Eqs.~(\ref{eq:alpha7}--\ref{eq:alpha9}) in the angular
distribution may allow this within a single decay mode.

We continue with the application of
Eq.~(\ref{eq:helicity-twoandthreebody}) to the case of
$J_1=J_2=1$ and $P_2=+1$ (e.g. $B\to \phi K_1$, $\phi\to
K\Kbar$, $K_1\to K\pi\pi$) and obtain
\begin{eqnarray}
{{16\pi} \over {9\Gamma}}~\!{d^3\Gamma \over
d{\cos\theta}_1 d{\cos\theta}_2d\Phi }
=  \alpha_1\times \,\cos^2\theta_1\, \sin^2\theta_2 \non\\
+\,\alpha_2\times \, {1\over 4}\,\sin^2\theta_1\,(1+\cos^2\theta_2) \non \\
-\,\alpha_3\times \, {1\over
 4}\,\sin^2\theta_1\,\sin^2\theta_2 \,\cos2\Phi \non \\
+\,\alpha_4\times \, {1\over 2}\,\sin^2\theta_1\,\sin^2\theta_2 \,\sin2\Phi \non \\
-\,\alpha_5\times \, {1\over 2\sqrt{2}}\,\sin2\theta_1\,\sin2\theta_2\,\cos\Phi \non \\
+\,\alpha_6\times \, {1\over 2\sqrt{2}}\,\sin2\theta_1\,\sin2\theta_2\,\sin\Phi \non \\
+\,\alpha_7\times \, r_1\,\sin^2\theta_1\,\cos\theta_2  \non \\
+\,\alpha_8\times \, r_1\,{1\over \sqrt{2}}\,\sin2\theta_1\,\sin\theta_2\,\sin\Phi \non \\
-\,\alpha_9\times \, r_1\,{1\over \sqrt{2}}\,\sin2\theta_1\,\sin\theta_2\,\cos\Phi \, .
\label{eq:va-angular}
\end{eqnarray}

Next, we apply Eq.~(\ref{eq:helicity-twoandthreebody}) to
the case of $J_1=1$, $J_2=2$, and $P_2=-1$ 
(e.g. $B\to \phi K_2$, $\phi\to K\Kbar$, $K_2\to K\pi\pi$) 
and obtain
\begin{widetext}
\begin{eqnarray}
{{64\pi(1+r_{02})} \over {45\Gamma}}~\!{d^3\Gamma \over
d{\cos\theta}_1 d{\cos\theta}_2d\Phi }
=  \alpha_1\times \,\{ \,\cos^2\theta_1\, \sin^4\theta_2\,
   + r_{02}\, {2\over3}\,\cos^2\theta_1\,\,(3\,\cos^2\theta_2\,-\,1)^2\,\}
\non \\
+ \alpha_2\times \,\{ \,{1\over3}\,\sin^2\theta_1\, \sin^2\theta_2\,(\,1\, +\, \cos^2\theta_2\,)
   + r_{02} \,\,{1\over2}\, \sin^2\theta_1\,\sin^2\,2\theta_2 \}
 \non \\
- \alpha_3\times \, \{ \,{1\over3}\,\sin^2\theta_1\, \sin^4\theta_2\,
   - r_{02} \,\,{1\over2}\, \sin^2\theta_1\,\sin^2\,2\theta_2 \}\, \cos\,2\Phi
 \non \\
+ \alpha_4\times \, \{ \,{2\over3}\,\sin^2\theta_1\, \sin^4\theta_2\,
   - r_{02} \,\, \sin^2\theta_1\,\sin^2\,2\theta_2 \}\, \sin\,2\Phi
 \non \\
- \alpha_5\times \, \{ \,\sqrt{2\over3}\,\sin\,2\theta_1\, \sin^3\theta_2\, \cos\theta_2\,
   - r_{02} \,\,\sqrt{{2\over3}}\, \sin\,2\theta_1\,\sin\,2\theta_2\,(\,3\cos^2\theta_2\,-\,1) \}\, \cos\,\Phi
 \non 
\end{eqnarray}
\begin{eqnarray}
+ \alpha_6\times \, \{ \,\sqrt{2\over3}\,\sin\,2\theta_1\, \sin^3\theta_2\, \cos\theta_2\,
   - r_{02} \,\,\sqrt{{2\over3}}\, \sin\,2\theta_1\,\sin\,2\theta_2\,(\,3\cos^2\theta_2\,-\,1) \}\, \sin\,\Phi
 \non \\
+ \alpha_7\times \, r_2  \,{4\over 3}\,\sin^2\theta_1\, \sin^2\theta_2\,\cos\theta_2
 \non \\
+ \alpha_8\times \, r_2 \,\sqrt{2\over3}\,\sin\,2\theta_1\, \sin^3\theta_2\, \sin\,\Phi
 \non \\
- \alpha_9\times \, r_2  \,\sqrt{2\over3}\,\sin\,2\theta_1\, \sin^3\theta_2\, \cos\,\Phi
\label{eq:vta-angular}
\end{eqnarray}
\end{widetext}

The $r_{02}$ and $r_2$ parameters in
Eq.~(\ref{eq:vta-angular}), just like $r_1$ in
Eqs.~(\ref{eq:vt-3body-angular}) and~(\ref{eq:va-angular}),
are a priori unknown.  However, if the dynamics of the
$X_2\to P_cP_dP_e$ decay is known, these parameters could be
further constrained, for example for a sequential two-body
decay chain such as $X_2\to P_cX_3$ with $X_3\to P_dP_e$.

Finally, we study the angular formalism of a spinless
particle decay $B\to X_1 X_2$, where both particles decay to
three pseudoscalars $X_1\to P_aP_bP_c$ and $X_2\to
P_dP_eP_f$.  An example of such a decay is $B\to\omega
K_{\sss J}$ or $\omega\omega$ with $\omega\to\pi^+\pi^-\pi^0$
and $K_{\sss J}\to K\pi\pi$.  Again, following the three-body
decay formalism \cite{berman1965}, we obtain
\begin{widetext}
\begin{eqnarray}
\label{eq:helicity-threebody}
{1\over \Gamma}
{d^3\Gamma \over d{\cos\theta}_1 d{\cos\theta}_2d\Phi} \propto
{1\over \sum|A_{\lambda}|^2}
\sum_{m}^{}\,\sum_{m^\prime}^{}\, |R_m|^2\, \,|R^\prime_{m^\prime}|^2\, 
\left|~\sum_{\lambda}^{} A_{\lambda} \, {\rm exp}({i\lambda\Phi})\, 
{d^{J_1\,}_{-\lambda,m}}(\pi-\theta_1)\,
{d^{J_2\,}_{\lambda,m^\prime}}(\theta_2)~\right|^2 \, .
\end{eqnarray}
\end{widetext}
where $|R_m|^2$ and $|R^\prime_{m^\prime}|^2$ are the
phenomenological parameters for $X_1\to P_aP_bP_c$ and
$X_2\to P_dP_eP_f$, respectively, with $2J_1+1$ values of $m$
and $2J_2+1$ values of $m^\prime$, as discussed with
reference to Eq.~(\ref{eq:helicity-twoandthreebody}).  The
same parity-conservation rules apply: $P_1=(-1)^{m+1}$ and
$P_2=(-1)^{m^\prime+1}$.  Eq.~(\ref{eq:helicity-threebody})
reduces to Eq.~(\ref{eq:helicity-twoandthreebody}) for the
decays with $J_1^{P_1}=1^-$, such as $B\to\omega
K^{(*)}_{\sss J}$ and $\omega\omega$.  The former is
described by
Eqs.~(\ref{eq:vt-3body-angular}--\ref{eq:vta-angular}) for
$K^*_2$, $K_1$, and $K_2\to K\pi\pi$, and the latter by
Eq.~(\ref{eq:vv-angular}).

The above angular formalism should facilitate experimental
analysis and measurements of $f_{\sss L}$, $f_{\sss \perp}$,
$\phi_{\sss \parallel}$, and $\phi_{\sss \perp}$ in various
$B\to VT$ and $VA$ decays.

The above results are useful in the analysis of the work by
Chen and Geng \cite{ChenGeng}, who proposed an explanation of
the $B\to V_1V_2$ and $B \to VT$ polarization results.
Briefly, it goes as follows. Consider the decay $\bd \to \phi
K^{*0}$. In addition to the naive SM contribution to $\bd \to
\phi K^{*0}$, Chen and Geng consider annihilation from the
operator ${\bar b} {\cal O} d {\bar d} {\cal O} s$, which
appears in the effective Hamiltonian [Eq.~(\ref{Heff})].
They use generalized factorization, which implies that this
term is factorizable; we denote it as $FA'$.  If $FA'$ is
sizeable, one has to consider $P'_{tc}$-$FA'$ interference in
computing $\fL$, $\fT$, etc.  Chen and Geng find
\beq
|A_{\sss L}(\phi K^*)|^2 \propto 1 + C_{\sss CG} (m_\phi^2 -
 m_{\sss K^*}^2) ~.
\label{chengeng}
\eeq
Here the first term is due to $P'_{tc}$, while the second is
due to $P'_{tc}$-$FA'$ interference. $C_{\sss CG}$ depends,
among other things, on the $\bd \to K^{*0}$ form factors.
These form factors, and $C_{\sss CG}$, are fixed by the
measured value of $\fL$ in $\bd \to \phi K^{*0}$.

The expression for $|A_{\sss L}(\phi K_2^*)|^2$ is identical
to that above, with the substitution $K^* \to K_2^*$. The key
point is that $(m_\phi^2 - m_{\sss K^*}^2)$ and $(m_\phi^2 -
m_{\sss K_2^*}^2)$ have opposite signs. Thus, assuming that
the $\bd \to K_2^{*0}$ form factors have the right size, if
$\fL$ is small in $\bd \to \phi K^{*0}$, it will be large in
$\bd \to \phi K_2^{*0}$, in agreement with observation. (The
explanation for $B^+ \to \phi K^{*+}$ is similar; the
operator ${\bar b} {\cal O} u {\bar u} {\cal O} s$
contributes here.) Chen and Geng assume that the time-like
form factor in $\bd \to \phi K^{*0}$ is related to the form
factor in $\bd \to \phi K_2^{*0}$, with the result that the
$B\to V_1V_2$ and $B \to VT$ polarization results are
reproduced.

Now, there can be objections to this explanation.  First, all
annihilation effects are thought to be small, and there is no
experimental evidence for a large $FA'$. Thus, the suggestion
that $FA'$ could be sizeable is somewhat arbitrary. Second,
Chen and Geng consider only one type of annihilation,
neglecting a second type which is also found in QCDf.  (It is
questionable whether this second type of annihilation
amplitudes is negligible compared to the first one.) Third,
the polarization results depend on (unknown) form
factors. Chen and Geng assume values for these form factors
which work, but this might not be the case.

However, putting aside these objections, the question is: can
we test the Chen-Geng explanation? Referring to
Eq.~(\ref{chengeng}), the reader could propose the following
possibility: it might be useful to consider decays in which
the two final-state particles have the same mass. Examples
include $\bs\to K^{\ast}\bar K^{\ast}$ and $\bs\to \phi\phi$.
In this case, $P'_{tc}$-$FA'$ interference apparently
vanishes, so that $\fTfL$ is small.  This can be tested.
Unfortunately, this idea does not work.  In
Eq.~(\ref{chengeng}), $C_{\sss CG}$ includes in the
denominator $m_q - m_{q_{sp}}$, where the decay is dominated
by the ${\bar b} \to {\bar q}$ penguin amplitude, and
$q_{sp}$ is the spectator quark.  In the above decays, this
difference vanishes ($q=s$). Thus, $P'_{tc}$-$FA'$
interference is of the form $0/0$.  After careful evaluation,
it is found that this interference is in fact nonzero.
 
Therefore, $\fTfL$ is arbitrary (it depends on unknown form
factors).  This same argument applies to all decays in which
the two final-state particles have the same mass. Thus, there
are no decay modes to test the Chen-Geng explanation. The
only thing to do is to compute the annihilation form factors
in $B\to VT$ decays, $\langle VT|O_i|0\rangle$, to see if
they agree with the Chen-Geng estimates.

In addition Chen, Geng and collaborators have considered
annihilation contributions in $B \to VA$ decays, $\langle
VA|O_i|0\rangle$. Assuming these contributions to be
factorizable, they discussed the decay $B \to \phi K_1$
\cite{ChenGeng2}. In their scenario, they conclude that the
annihilation contributions can be neglected, which implies
that $\fTfL$ in $B \to \phi K_1$ is small. However, their
conclusion assumes that the $B \to VA$ annihilation form
factors are similar to those of $B \to V_1V_2$.
Unfortunately, the form factors relevant to $VA$ final states
are unknown -- if one takes any values for them, $\fTfL$ in
$B \to \phi K_1$ can be large or small. Thus, this is not a
real test of the Chen-Geng explanation.

Penguin annihilation and rescattering make no prediction for
the polarization in $B \to \phi K_1$, since this is a
different final state from $\phi K^*$. NP also makes no
prediction, since its result depends on the unknown $B \to A$
form factors. Despite the fact that there is no firm
prediction for the polarization in $B \to \phi K_1$ -- or
perhaps because of it -- this is an important measurement to
make.  Now, $K_1$ decays only to 3 bodies, e.g.\ to $K
\pi\pi$.  Thus, the angular analysis of the 3-body decay
described above will be necessary to get helicity
information.

\section{Conclusions}

In $B\to V_1V_2$ decays ($V_i$ is a light charmless vector
meson), the final-state particles can have transverse or
longitudinal polarization.  Within the standard model (SM),
the naive expectation is that the fraction of
transverse-polarization decays is much less than the fraction
of longitudinal decays: $\fTfL \ll 1$. However, it was found
in $B\to\phi K^*$ that $\fTfL \simeq 1$. This is the
``polarization puzzle.'' It is not necessary to invoke new
physics (NP), though one can.  There are two extensions
beyond the naive SM in which the polarization puzzle is
explained: penguin annihilation and rescattering.  One can
also look at $B \to VT$ decays ($T$ is a tensor meson). In
$B\to\phi K_2^*$ $\fTfL$ is observed to be small.  In this
paper, we look at the polarizations in $B \to VT$ decays,
both in the SM (naive and extended) and with NP. The idea is
to examine the prediction for polarization in $B \to VT$, to
see if it is in agreement with the measurement.

We begin by considering the naive prediction of the SM. It
depends on the $B \to T$ form factors, which are unknown at
present. These form factors can be expressed in terms of two
independent, universal form factors,
$\widetilde{\zeta}_\perp$ and $\widetilde{\zeta}_\|$. We
consider $\widetilde{\zeta}_\perp \approx
\widetilde{\zeta}_\|$ [case (a)] and $\widetilde{\zeta}_\perp
\ll \widetilde{\zeta}_\|$ with $\widetilde{\zeta}_\| \sim
({m_{\sss B} / m_{\sss V}})\widetilde{\zeta}_\perp $ [case
(b)]. In both cases we find that $\fTfL(B\to\phi K_2^*) \ll
1$. We therefore conclude that the naive SM reproduces the
polarization measurement in $B\to\phi K_2^*$.

The polarization predictions of both penguin annihilation and
rescattering are not certain. That is, the predictions depend
on a new set of parameters for each final state. Thus, the
final-state polarization in $B\to\phi K_2^*$ can be large or
small. It is therefore possible that both explanations agree
with the $\fTfL$ measurements in $B\to\phi K^*$ and $B\to\phi
K_2^*$.

The combined branching-ratio and CP-asymmetry measurements in
$B\to\pi K$ decays are also in disagreement with the SM
predictions (the ``$K \pi$ puzzle''). In Ref.~\cite{NP} it
was found that only two new-physics operators can account
for the discrepancies in both the $\pi K$ data and the $\phi(
\rho) K^*$ polarization measurements. We examine the
predictions of these operators for polarization in $B\to\phi
K_2^*$. We find a dependence on the $B \to T$ form factors.
If they obey case (a), then NP cannot explain the $B\to\phi
K_2^*$ polarization data.  However, if they are described by
case (b), then NP can account for the measurements in
$B\to\pi K$, $B\to\phi K^*$ and $B\to\phi K_2^*$. This can be
tested by explicit computations of the $B \to T$ form
factors.

Finally, most of the polarization measurements to date use
the two-body decays of particles.  In this paper, we present
the general angular analysis. In particular, we show how to
get helicity information using three-body decays.  This is
important for vector, tensor, and axial-vector final-state
mesons.

\bigskip
\noindent
{\bf Acknowledgments}:
This work was financially supported by NSERC of Canada (DL,
MN \& AS), and the U.S. NSF (YG \& AG) and A.~P.~Sloan
Foundation (AG).


\bibliographystyle{h-physrev2-original}   

\end{document}